\documentclass[letterpaper,12pt,titlepage,oneside,final]{book}

\newif\ifindustry

\industryfalse

\usepackage{multirow}
\usepackage[british,UKenglish]{babel}  
\usepackage[letterpaper,margin=1in]{geometry} 
\usepackage{natbib} 
\usepackage{graphicx} 
\usepackage[export]{adjustbox}
\usepackage[absolute,overlay]{textpos}
\usepackage{tikz}
\usepackage[linesnumbered,lined,boxed,commentsnumbered]{algorithm2e}

\usepackage{amsfonts,amssymb}	
\usepackage{mathrsfs} 
\usepackage{amsmath}
\usepackage{lastpage}
\usepackage[bookmarks=true, colorlinks=false]{hyperref} 
\DeclareGraphicsExtensions{.pdf,.png,.jpg}
\graphicspath{{./figures/}} 
\usepackage{mathtools}
\usepackage{setspace}
\usepackage{bbm}

\usepackage{listings}
\usepackage{xcolor}

\usepackage{booktabs, multicol, multirow,xcolor,colortbl,threeparttable}

\providecommand{\tightlist}{%
  \setlength{\itemsep}{0pt}\setlength{\parskip}{0pt}}
  
\usepackage[scaled=0.90]{helvet} 
\usepackage{courier} 
\normalfont
\usepackage[T1]{fontenc}

\newcommand{\mytype}{ActSc 906 \& 972 Final Project}
\newcommand{\mystream}{Master of Quantitative Finance}

\newcommand{\myname}{Jinyang Li}

\newcommand{\mytitle}{Valuing American Options by Simulation: Least Square and Machine learning Approaches}

\hypersetup{
 pdfauthor={\myname},
 pdftitle={\mytitle},
 pdfsubject={\mytype},
 pdfkeywords={\mystream}
}

\begin{document}

\pagestyle{empty}
\pagenumbering{roman}

\begin{titlepage}
    \begin{center}
        \vspace*{1.0cm}

        \Large
        {\bf A DEEP REINFORCEMENT LEARNING FRAMEWORK FOR FINANCIAL PORTFOLIO MANAGEMENT}

        \vspace*{1.3cm}

        \normalsize
        by \\

        \vspace*{0.6cm}

        \Large
        Jinyang Li \\

        \vspace*{3.0cm}

        \normalsize
        December 2019 \\
        
        \vspace*{0.8cm}

        \normalsize
        Supervisor: David Saunders \\
        
        \vspace*{4.0cm}

        \normalsize
        A Master research paper \\
        presented to the University of Waterloo \\ 
        in fulfillment of the \\
        thesis requirement for the degree of \\
        Master of Quantitative Finance \\

        \vspace*{2.0cm}
    \end{center}
\end{titlepage}

\pagestyle{plain}
\setcounter{page}{2}

\cleardoublepage 


\begin{center}\textbf{Abstract}\end{center}

In this research paper, we investigate into a paper named ``A Deep Reinforcement Learning Framework for the Financial Portfolio Management Problem''. It is a portfolio management problem which is solved by deep learning techniques. The original paper proposes a financial-model-free reinforcement learning framework, which consists of the Ensemble of Identical Independent Evaluators (EIIE) topology, a Portfolio-Vector Memory (PVM), an Online Stochastic Batch Learning (OSBL) scheme, and a fully exploiting and explicit reward function. Three different instants are used to realize this framework, namely a Convolutional Neural Network (CNN), a basic Recurrent Neural Network (RNN), and a Long Short-Term Memory (LSTM). The performance is then examined by comparing to a number of recently reviewed or published portfolio-selection strategies. We have successfully replicated their implementations and evaluations. Besides, we further apply this framework in the stock market, instead of the cryptocurrency market that the original paper uses. The experiment in the cryptocurrency market is consistent with the original paper, which achieve superior returns. But it doesn't perform as well when applied in the stock market.

\cleardoublepage


\begin{center}\textbf{Acknowledgements}\end{center}

I would first like to thank my supervisor Prof. David Saunders of the Faculty of Mathematics at University of Waterloo. The door to his office was always open whenever I ran into a trouble spot or had a question about my research or writing. He consistently allowed this paper to be my own work, but steered me in the right the direction whenever he thought I needed it. His guidance helped me in all the time of research and writing of this Master's research paper. I could not have imagined having a better advisor and mentor for my Master's study.

My sincere thanks also goes to Zhengyao Jiang, who is the author of the original paper that I have studied in this research paper, provided me an opportunity to discuss my confusions and offered great help to me during my implementation. Without his precious support it would not be possible to conduct this research.

Last but not the least, none of this could have happened without my family. My parents have paid the subscription fee for me to get the stock market data. They also offered their encouragement through phone calls every week – despite my own limited devotion to correspondence. My wife was always on my side and encouraged me whenever I felt frustrated. This Master's research paper stands as a testament to your unconditional love and encouragement.

\cleardoublepage

\renewcommand\contentsname{Table of Contents}
\tableofcontents
\cleardoublepage
\phantomsection    

\addcontentsline{toc}{chapter}{List of Tables}
\listoftables
\cleardoublepage
\phantomsection		

\addcontentsline{toc}{chapter}{List of Figures}
\listoffigures
\cleardoublepage
\phantomsection		

\pagenumbering{arabic}


\clearpage

\doublespacing

\pagestyle{plain}
\pagenumbering{arabic}



\chapter{Introduction}
\label{chp:Introduction}

The portfolio management problem is investigated in this research paper and we focus on a cutting-edge paper using deep machine learning techniques to solve it which is written by \cite{jiang2017deep}, here after referred to as ``the original paper''. As we know, it is an optimization problem that continuously reallocates capital into different financial products, according to certain objective. The objective typically maximizes factors such as expect return, and minimizes the risk such as the standard deviation. The portfolio can be of wide range from tangible assets to intangible assets.

The original paper proposes a reinforcement learning network specially designed for the task of portfolio management. The core of the framework is the Ensemble of Identical Independent Evaluators (EIIE) topology. An Identical Independent Evaluators (IIE) is a neural network whose job is to analyze the history of an asset and then evaluates its potential growth for the immediate future. Three different types of IIEs are tested in the original paper, namely a Convolutional Neural Network (CNN), a basic Recurrent Neural Network (RNN) and a Long Short Term Memory (LSTM). The IIEs are trained on each asset and then assembled to a softmax layer for the new portfolio weights for the coming trading period. In order to take into account the effect of transaction costs, the historical portfolio weights are also inputs to the EIIE. For this purpose, the portfolio weights of each period are recored in a Portfolio Vector Memory (PVM). The EIIE is trained in an Online Stochastic Batch Learning (OSBL) scheme, which is compatible with both pre-trade training and online training during back-test or online trading. 

This framework is not restricted to any particular market. The original paper tests its validity and profitability in the cryptocurrency exchange market. In this paper, we first implemented the experiments based on the same setting of Jiang's paper. Then, we further evaluate this framework in the stock market. For both experiments, the performance of the three EIIEs (CNN, RNN and LSTM) is compared with some recently published or reviewed portfolio selection strategies \citep{li2016olps,li2014online}. The EIIEs significantly beat all the other strategies when it is in the cryptocurrency market, but the performance is average in stock market.

The remainder of this paper is organized as follows. \textbf{Chapter 2} presents background information and related works in this field. \textbf{Chapter 3} defines the portfolio management problem that the framework is aiming to solve. \textbf{Chapter 4} is the implementation specifications, introducing the asset pre-selection process and the input price tensor. \textbf{Chapter 5} presents the reinforcement learning framework. The results of two experiments under the cryptocurrency and stock markets are illustrated in \textbf{Chapter 6}, as well as the evaluation and discussion. Finally, \textbf{Section 7} concludes the whole project and lists possible future works.

\chapter{Background and Related Work}
\label{sec:rw}

In this chapter, we first talk about the background of portfolio management theory and the cryptocurrency market. Then we present related works that have been done using machine learning techniques.

\section{Background}
\label{sec:Background}

\subsection{Portfolio Management Theory}
\label{subsec:Portfolio Management Theory}

Modern Portfolio Theory (MPT) was first introduced by 
\citeauthor{markowitz1952portfolio}, who was later awarded a Nobel price for developing this theory. In a series of papers \citep{markowitz1952portfolio,markowitz1959portfolio}, the Markowitz model was proposed and developed. The model, also known as the mean-variance model, assists in the selection of the most efficient portfolio by analyzing various possible portfolios of given securities. It uses statistical analysis for measurement of risk and mathematical programming for selection of assets in an efficient manner. According to the model, it is possible to construct the efficient frontier of optimal portfolios offering the maximum possible expected return for a given level of risk. 

Another type of theory called Capital Growth Theory (CGT) was proposed shortly afterwards by \cite{kelly1956new,hakansson1995capital}, which is primarily originated from information theory. Instead of focusing on a single-period portfolio selection as it is in the MPT, CGT focuses on multiple-period or sequential portfolio selection, aiming to maximize the portfolio's expected growth rate or expected log return. While both theories solve the task of portfolio selection, the latter is fitted to the online scenario, which naturally consists of multiple periods and is the basis for the paper that we focus on. 

Online portfolio selection, which sequentially selects a portfolio over a set of assets in order to achieve certain targets, is a natural and important task for asset portfolio management. Aiming to maximize the cumulative wealth, several categories of algorithms have been proposed to solve this task. According to \cite{li2014online}, online portfolio selection algorithms can be classified into four categories, namely ``Follow-the-Winner'', ``Follow-the-Loser'', ``Pattern-Matching Approaches'' and ``Meta-Learning Algorithms''.

\begin{itemize}
    \item \textbf{Follow-the-Winner}: tries to asymptotically achieve the same growth rate as that of an optimal strategy, which is often based on the Capital Growth Theory.
    \item \textbf{Follow-the-Loser}: transfers the wealth from winning assets to losers, which seems contradictory to common sense but empirically often achieves significantly better performance. 
    \item \textbf{Pattern-Matching Approaches}: tries to predict the next market distribution based on a sample of historical data and explicitly optimizes the portfolio based on the sampled distribution.
    \item \textbf{Meta-Learning Algorithms}: combines multiple strategies from the above three categories.
\end{itemize}

We selected some algorithms from the above categories, together with three benchmarks namely Best stock, Uniformly Constant Rebalanced Portfolios and Buy And Hold, to evaluate the performance of the EIIE neural networks. A table summarizing those algorithms is as below, see Table \ref{tab:OnlinePortfolioSelection}.

\begin{table}[htbp]

  \centering

  \setlength\tabcolsep{3pt} 
  \footnotesize
    \begin{tabular}{cll}
    \toprule
    \textbf{Classifications } & \textbf{Algorithms } & \textbf{Abbreviation} \\
    \midrule
    \multirow{3}[2]{*}{Benchmarks } & Buy And Hold & ubah \\
          & Best Stock & best \\
          & Uniformly Constant Rebalanced Portfolios  & ucrp \\
    \midrule
    \multirow{3}[2]{*}{Follow-the-Winner } & Universal Portfolios & up \\
          & Exponential Gradient & eg \\
          & Online Newton Step & ons \\
    \midrule
    \multirow{6}[2]{*}{Follow-the-Loser } & Anti Correlation & anticor \\
          & Online Moving Average Reversion & olmar \\
          & Passive Aggressive Mean Reversion & pamr \\
          & Weighted Moving Average Passive Aggressive & wmamr \\
          & Confidence Weighted Mean Reversion & cwmr \\
          & Robust Median Reversion & rmr \\
    \midrule
    \multirow{3}[2]{*}{Pattern-Matching} & Nonparametric Histogram Log-optimal & bk \\
          & Nonparametric Nearest Neighbor Log-optimal & bnn \\
          & Correlation-driven Nonparametric Learning & cornk \\
    \midrule
    Others & Constant Rebalanced Portfolio & m0 \\
    \bottomrule
    \end{tabular}%
 \caption{General classification for the state-of-the-art online portfolio selection algorithms.}

  \label{tab:OnlinePortfolioSelection}%
\end{table}%

\subsection{Cryptocurrencies}
\label{subsec:Cryptocurrencies}

The original paper tests the framework in the cryptocurrency market. A cryptocurrency is a digital asset designed to work as a medium of exchange that uses strong cryptography to secure financial transactions, control the creation of additional units, and verify the transfer of assets \citep{narayanan2016bitcoin,chohan2017cryptocurrencies,schueffel2017concise}. Cryptocurrencies use decentralized control as opposed to centralized digital currency and central banking systems \citep{szabo2015if}. 

Bitcoin, first released as open-source software in 2009, is generally considered the first decentralized cryptocurrency. Since the release of bitcoin, over 6,000 altcoins, which are alternative variants of bitcoin, or other cryptocurrencies, have been created \citep{sagona2015bitcoin}. The motivation of inventing so many variants is that there are flaws in Bitcoin and those invariants are invented to overcome these defects hoping their inventions will eventually replace Bitcoin. However, as of December 2019, Bitcoin takes 66.6\% of the total market capital of all cryptocurrencies, which is nearly 198 billions in USD\footnote{Crypto-currency market capitalizations, http://coinmarketcap.com/, accessed: 2019-12-14.}. Therefore, Bitcoin is still the dominant cryptocurrency in the market and many other currencies can only be traded against Bitcoin. As a result, the original paper uses Bitcoin to be the unit when measuring the price of each cryptocurrency.

\subsection{Reinforcement Learning}
\label{subsec:reinforcement learning}

Reinforcement learning is an area of machine learning concerned with how software agents ought to take actions in an environment in order to maximize some notion of cumulative reward. Reinforcement learning is one of three basic machine learning paradigms, alongside supervised learning and unsupervised learning.

It differs from supervised learning in not needing labelled input/output pairs to be presented, and in not needing sub-optimal actions to be explicitly corrected \citep{kaelbling1996reinforcement}. Instead the focus is on finding a balance between exploration (of uncharted territory) and exploitation (of current knowledge).

\begin{figure}[htbp]
    \centering
    \includegraphics[width=0.7\textwidth]{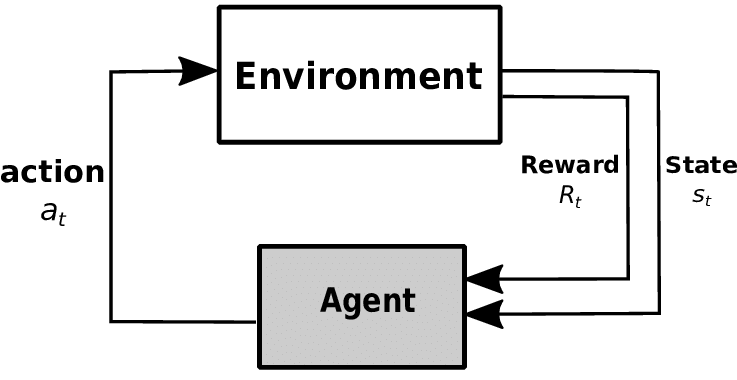}
    \caption{The Typical Framing of a Reinforcement Learning (RL) scenario}
    \label{fig:RL_Framing}
\end{figure}

Reinforcement learning involves an agent and an environment: every time step, the agent chooses an action, and the environment will return a reward and transitions into the next state. A typical framework is shown in Figure \ref{fig:RL_Framing}. The agent's interaction with the environment is broken up into a series of episodes, each consisting of one or more time steps/actions. The process continues until a terminal state is reached, at which point the episode ends.

\section{Related Work}
\label{sec:Related Work}

Many machine learning approaches have been proposed for financial market trading. Most of them focus on stock price prediction or trends prediction, see \cite{heaton2017deep,niaki2013forecasting,freitas2009prediction}. The neural networks take the historical prices or indicators as input and then output the stock price or trend for the next period. Depending on the prediction, certain trading action will be taken, such as buy, sell or hold. Those methods are straightforward and most are supervised learning, or essentially regression problems. The portfolio performance largely depends on the prediction accuracy, but it turns out the future market prices are difficult to predict. Moreover, there is still a layer between the neural network's output and trading action, which needs extra logic that can be based on human decisions. In this case, the whole approach is not purely machine learning and thus not very extendable or adaptable.

Using Reinforcement learning is a successful attempt when trying to apply model-free and fully machine learning schemes to algorithmic trading problems, without predicting future prices. Related work includes: \cite{moody2001learning,dempster2006automated,cumming2015investigation,deng2016deep}. However, those algorithms consider only the one asset case and are not applicable when it comes to portfolio management problems, where trading agents manage multiple assets.

One major problem with many reinforcement learning problem is the discrete action spaces that cannot be applied directly to the portfolio management problem, where actions are continuous. The market actions can be discretized but this comes with unknown risk. For example, the agent may invest all the capital in one asset, without spreading the risk to the rest of the market. As a result, instead of discrete action spaces, the agent in the original paper outputs a vector of portfolio weights as next action which is continuous. 
\chapter{Problem Definition}
\label{chp:Problem Definition}

This chapter defines the mathematical setting of the portfolio management problem. We first define the time discretization, then formulate the problem using mathematical symbols. Afterwards, transaction costs will be introduced as well as the two assumptions we imposed on our framework.

\section{Two Assumptions}
\label{sec:Two assumptions}

The validation of the framework is based on two assumptions.
\begin{enumerate}
    \item Zero slippage: we assume that the volume of the market is high enough that each order can be filled with the market price at placement. 
    \item Zero money impact: the capital of the portfolio is small enough that it has no influence on the market.
\end{enumerate}

These are the reason that later we select cryptocurrencies/stocks with high trading volume, in which case the above two assumptions that the framework is based on will be satisfied.

\section{Mathematical Formalism}
\label{sec:Mathematical Formalism}

\begin{figure}[htbp]
    \centering
    \includegraphics[width=\textwidth]{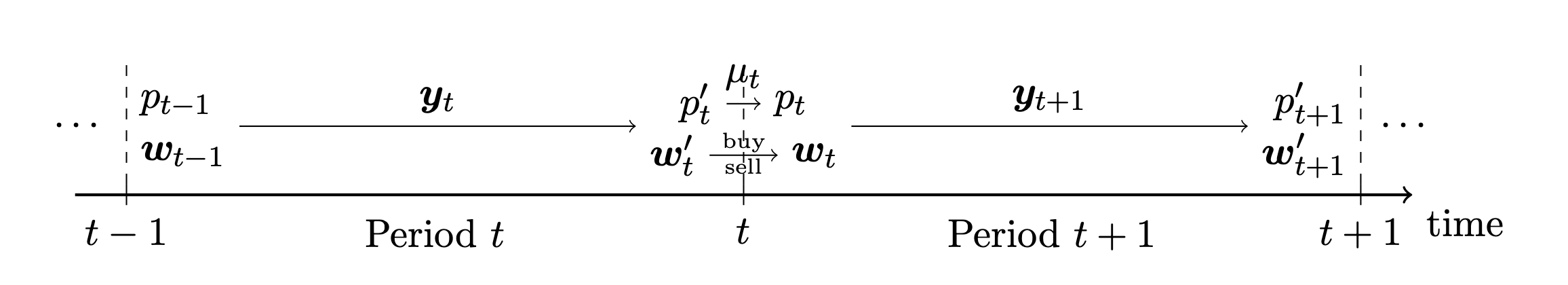}
    \caption{Illustration of trading between periods.}
    \label{fig:Trade_periods_illustration}
\end{figure}

The whole trading process is divided into periods of equal length $T$, which is set to 30 mins in our later experiments. Shown in Figure \ref{fig:Trade_periods_illustration}, period $t$ and $t+1$ are two consecutive periods. In each period, the price movement can be described using four features, namely opening price, closing price, highest price and lowest price. By the zero slippage assumption in Section \ref{sec:Two assumptions}, the closing price of perod $t$ is the opening price of period $t+1$.

Considering a portfolio of $m$ assets, we define the $price\ vector$ as follows:
\begin{itemize}
    \item $\boldsymbol{v_t}=(v_{0,t},v_{1,t},\dots,v_{m,t})$ is the closing prices of the $t$th period.
    \item $\boldsymbol{v_t}^{(hi)}=(v^{(hi)}_{0,t},v^{(hi)}_{1,t},\dots,v^{(hi)}_{m,t})$ is the highest prices of the $t$th period.
    \item $\boldsymbol{v_t}^{(lo)}=(v^{(lo)}_{0,t},v^{(lo)}_{1,t},\dots,v^{(lo)}_{m,t})$ is the lowest prices of the $t$th period.
\end{itemize}

We use Bitcoin as the cash and all other prices are quoted in cash, therefore the first element of each price vector will always be $1$, that is $v_{0,t}=v^{(hi)}_{0,t}=v^{(lo)}_{1,t}=1, \forall t$.

We then define the $relative\ price\ vector$ of the $t$th trading period $y_t$ as:
\begin{equation}
    \boldsymbol{y_t} \coloneqq \boldsymbol{v_t} \oslash \boldsymbol{v_{t-1}}
\end{equation}
which is the element-wise division of $v_t$ by $v_{t-1}$. The elements of $y_t$ are the quotients of closing prices and opening prices for individual asset in the period. The relative price vector can be used to calculate the change in total portfolio value in a period. 

Let $p_{t-1}$ denote the portfolio value at the beginning of period $t$, then we have,
\begin{equation}
    p_t=p_{t-1} \boldsymbol{y_t}\cdot \boldsymbol{w_{t-1}}
\end{equation}
where $\boldsymbol{w_{t-1}}$ is the portfolio weight vector (hereinafter referred to as the $portfolio\ vector$) at the beginning of period $t$. By definition, all elements in the $portfolio\ vector$ sum up to 1. Then the $rate\ of\ return$ for period $t$ is

\begin{equation}
    \rho_t \coloneqq \frac{p_t}{p_{t-1}} -1 =\boldsymbol{y_t}\cdot \boldsymbol{w_{t-1}} -1
    \label{eq:rateofreturn}
\end{equation}
and the corresponding $corresponding\ logarithmic\ rate\ of\ return$ is 
\begin{equation}
    r_t \coloneqq \ln{\frac{p_t}{p_{t-1}}}  =\ln{\boldsymbol{y_t}\cdot \boldsymbol{w_{t-1}}}
    \label{eq:lograteofreturn}
\end{equation}

When there is no transaction cost, the final portfolio value will be
\begin{equation}
    p_f=p_0\exp{\sum_{t=1}^{t_f+1}r_t}=p_0\prod_{t=1}^{t_f+1}\boldsymbol{y_t}\cdot \boldsymbol{w_{t-1}} 
\end{equation}
where $p_0$ is the initial investment amount. The goal of our framework is to maximize the final portfolio value for a given time frame.

\section{Transaction Costs}
\label{sec:Transaction Costs}

In the real world financial market, transaction costs are not immaterial and may affect the portfolio significantly. Many strategies may have a great performance without transaction costs, but become terrible when they are included. Therefore, it is crucial for us to include transaction costs into our framework. In our framework, we assume that the costs mainly come from commission fee and we assume a constant commission rate using the recursive formula by \cite{ormos2013performance}.

The portfolio vector at the end of period $t$ and before any action is taken is denoted by $\boldsymbol{w_t}'$. Due to the price move in the period, the weights $\boldsymbol{w_t}'$ evolve into

\begin{equation}
      \boldsymbol{w_t'} = \frac{\boldsymbol{y_t} \odot \boldsymbol{w_{t-1}} }{\boldsymbol{y_t}\cdot \boldsymbol{w_{t-1}} }
\end{equation}
where $\odot$ is the element-wise multiplication. The task for the portfolio manager is to redistribute the whole portfolio such that the initial weights at the beginning of period $t+1$ are changed from $\boldsymbol{w_t'}$ to $\boldsymbol{w_t}$. As a result, a series of trading actions will take place at time $t+1$ as well as the involvement of commission fees. Suppose such redistribution action will shrink the portfolio value by a $transaction\ remainder\ factor$ $\mu_t$ where $\mu_t \in (0,1]$, the factor can be determined by

\begin{equation}
    \mu_t=\frac{p_t}{p_t'}
\end{equation}
where $p_{t}'$ is the portfolio value at the end of period $t$ before any action is taken and $p_t$ at the beginning of period $t+1$ after transactions costs. Under this situation, \eqref{eq:rateofreturn} and \eqref{eq:lograteofreturn} are now

\begin{equation}
    \rho_t \coloneqq \frac{p_t}{p_{t-1}} -1 =\frac{\mu_t p_t'}{p_{t-1}} -1=\mu_t\boldsymbol{y_t}\cdot \boldsymbol{w_{t-1}} -1
    \label{eq:rho_w_transaction}
\end{equation}

\begin{equation}
    r_t \coloneqq \ln{\frac{p_t}{p_{t-1}}}  =\ln{\mu_t\boldsymbol{y_t}\cdot \boldsymbol{w_{t-1}}}
    \label{eq:logreturn_w_transaction}
\end{equation}
and the final portfolio value becomes
\begin{equation}
        p_f=p_0\exp{\sum_{t=1}^{t_f+1}r_t}=p_0\prod_{t=1}^{t_f+1}\mu_t\boldsymbol{y_t}\cdot \boldsymbol{w_{t-1}}
        \label{eq:fpv}
\end{equation}
Once again, a graphical representation of the relationship among portfolio vectors can be found in Figure \ref{fig:Trade_periods_illustration}.

Let $c_s, c_p \in [0,1)$ denote the commission rate for selling and purchasing respectively, $\mu_t$ can be calculated by

\begin{equation}
    \mu_t=\frac{1}{1-c_p w_{t,0}}\left[1-c_p w'_{t,0}-(c_s+c_p-c_s c_p)\sum_{i=1}^{m}(w'_{t,i}-\mu_t w_{t,i})^+ \right]
    \label{eq:mu_t_eq}
\end{equation}
The proof of \eqref{eq:mu_t_eq} will be given in Appendix \ref{AppendixA}.
\newcommand{\defeq}{\vcentcolon=}
\let\oldnl\nl
\newcommand{\nonl}{\renewcommand{\nl}{\let\nl\oldnl}}
\chapter{Data Structure}
\label{chp:Algorithms}
Data is an important consideration before feeding into the neural network framework. Garbage in, garbage out. So we need a careful analysis on the treatment of data. In this chapter, we will go through how we select the assets and how the data are stored in the price tensor.

\section{Asset Selection}
\label{sec:AssetSelection}

In the original paper, the top 11 non-cash assets from the cryptocurrency market ranking by trading volume are selected to construct the portfolio. Together with the cash, Bitcoin, the size of the portfolio, $m+1$, is 12. As for the stock market experiment, the top 11 stocks are selected according to their volume as well.

The reasons for the above selection is as follows:
\begin{enumerate}
    \item Bigger volume implies better market liquidity of an asset, in turn meaning that our experiment satisfies the Assumption 1 in Section \ref{sec:Two assumptions}. It also suggests that the portfolio is less likely to have influence on the market, making the environment close fulfilling the Assumption 2.
    \item The market of cryptocurrency is not stable. Some cryptocurrencies may suddenly jump or drop in volume in a short time. Therefore, the volume is measured on a longer time-frame, where in the later experiments, 30-day average volumes are used.
\end{enumerate}

The whole time frame is split into training sets and test sets. The former is used for offline training of the policy network while the later is used for back-testing as well as online training. Choosing the top volume assets just before the end of the whole time frame can be subject to the survivor bias. The trading volume of an asset is correlated to its popularity, which in turn is governed by its historic performance. Given the future ranking by volume will inevitably pass future price information to the experiments. In order to avoid this bias as much as possible, the average volume is taken just before the beginning of the back-tests.

\section{Price Tensor}
\label{sec:pricetensor}

The data is fed into the neural network in the form of price tensors. The neural network will then generate the output of portfolio vectors. We will talk about the structure of the price tensor and its normalization scheme.

The price tensor $\boldsymbol{X_t}$ is the input to the neural networks at the end of period $t$, which has a dimension of $(f,n,m)$ where $m$ is the number of preselected non-cash assets, $n$ is the number of input periods before t, and $f$ is the feature number. The features are chosen to be the closing price, highest price and lowest price in a period. Using the notation in Section \ref{sec:Mathematical Formalism}, these are $v_{i,t},v_{i,t}^{(hi)},v_{i,t}^{(lo)}$ for asset $i$ on period $t$.

An important step before directly feeding the data into the networks is normalization. The absolute price is normalized by the latest closing price in order to reflect the changes in prices which actually determine the performance of the portfolio management. The price tensor $\boldsymbol{X_t}$ is defined as

\begin{align}
    \boldsymbol{X_T} &= (\boldsymbol{V_{t}},\boldsymbol{V_{t}^{(hi)}},\boldsymbol{V_{t}^{(lo)}})
\end{align}
where $\boldsymbol{V_{t}},\boldsymbol{V_{t}^{(hi)}},\boldsymbol{V_{t}^{(lo)}}$ are the normalized price matrices,
\begin{equation*}
\begin{split}
    \boldsymbol{V_{t}} &= [\boldsymbol{v_{t-n+1}} \oslash \boldsymbol{v_{t}}, \boldsymbol{v_{t-n+2}} \oslash \boldsymbol{v_{t}},\dots,\boldsymbol{v_{t-1}} \oslash \boldsymbol{v_{t}},\boldsymbol{1}]\\
    \boldsymbol{V_{t}^{(hi)}} &= [\boldsymbol{v_{t-n+1}^{(hi)}} \oslash \boldsymbol{v_{t}}, \boldsymbol{v_{t-n+2}^{(hi)}} \oslash \boldsymbol{v_{t}},\dots,\boldsymbol{v_{t-1}^{(hi)}} \oslash \boldsymbol{v_{t}},\boldsymbol{v_{t}^{(hi)}} \oslash \boldsymbol{v_{t}}]\\
    \boldsymbol{V_{t}^{(lo)}} &= [\boldsymbol{v_{t-n+1}^{(lo)}} \oslash \boldsymbol{v_{t}}, \boldsymbol{v_{t-n+2}^{(lo)}} \oslash \boldsymbol{v_{t}},\dots,\boldsymbol{v_{t-1}^{(lo)}} \oslash \boldsymbol{v_{t}},\boldsymbol{v_{t}^{(lo)}} \oslash \boldsymbol{v_{t}}]
\end{split}
\end{equation*}
with $\boldsymbol{1}=(1,\dots,1)^T$ and $\oslash$ being the element-wise division operator.

At the end of period $t$, the portfolio manager comes up with a portfolio vector $w_t$ using merely the information from the price tensor $\boldsymbol{X_t}$ and the previous portfolio vector $w_{t-1}$, according to some policy $\pi$. In other words, $w_t = \pi(\boldsymbol{X_t},w_{t-1})$. At the end of period $t + 1$, the logarithmic rate of return for the period due to decision $w_t$ can be calculated with the additional information from the price change vector $y_{t+1}$, using Equation \eqref{eq:logreturn_w_transaction}. In the language of reinforcement learning, $r_{t+1}$ is the immediate reward to the portfolio management agent for its action $w_t$ under the environment $\boldsymbol{X_t}$.

\chapter{Reinforcement Learning Framework}
\label{chp:RL Framework}
\newcommand\Tau{\mathcal{T}}
Extending what we have discussed in Subsection \ref{subsec:reinforcement learning}, we will investigate into how the original paper builds up the whole reinforcement learning framework. First, we will talk about how the portfolio problem defined before can be applied to reinforcement learning, specifically explains the representation of each element. Then, we will look into the policy network which is the core innovation from the original paper. Three different deep neural networks are used here to construct the policy function, namely CNN, RNN and LSTM. Three innovations will also be discussed, 
namely the network topologies, the portfolio-vector memory and the online stochastic batch learning scheme.

\section{Reinforcement Learning Elements}
\label{sec:RL elements}

As shown in Figure \ref{fig:RL_Framing}, several elements are crucial to the framing of a reinforcement learning setup, which will be explained in detail.

\subsubsection{The Agent and the Environment}
\label{subsubsec:The Agent and the Environment}

The agent in this portfolio management problem is the software portfolio manager performing trading-actions in the environment of a financial market. This environment is comprised of all available assets in the market and the expectations of all market participants towards them.

\subsubsection{The State and the Action}
\label{subsubsec:The State and the Action}
In theory, the state at time $t$ should be all the information up until $t$. However, it is impossible to get and store all such information considering such a large and complex environment. It is believed that all relevant information is reflected in the prices of the assets, which are already publicly available to the agent \citep{kirkpatrick2010technical,lo2000foundations}. Under this point of view, sub-sampling schemes for the order-history information are employed to be the state representation of the market environment. Besides, the action made at the beginning of period $t$ will affect the reward of period $t+1$, and as a result will affect the decision of its action. As a result, the state at $t$ is represented as the pair of $\boldsymbol{X_t}$ and $w_{t-1}$,
\begin{equation}
    \boldsymbol{s_t}=(\boldsymbol{X_t}, w_{t-1})
\end{equation}
The action $a_t$ at time $t$ is therefore
\begin{equation}
    \boldsymbol{a_t}=\boldsymbol{w_t}
\end{equation}
where $\boldsymbol{w_0} = (1,0,\dots,0)^T$.

\subsubsection{The Reward Function}
\label{subsubsec:The Reward Function}
The objective of the agent is to maximize the portfolio value $p_f$ of Equation \eqref{eq:fpv} at the end of $t_f+1$ period. Or equivalently, the job is to maximize the average logarithmic accumulated return $R$,

\begin{equation}
    R(\boldsymbol{s_1},\boldsymbol{a_1},\dots,\boldsymbol{s_{t_f}},\boldsymbol{a_{t_f}},\boldsymbol{s_{t_f+1}}) = \frac{1}{t_f}\ln{\frac{p_f}{p_0}}=\frac{1}{t_f}\sum_{t=1}^{t_f+1}\ln{\mu_t \boldsymbol{y_t}\cdot\boldsymbol{w_{t-1}}}=\frac{1}{t_f}\sum_{t=1}^{t_f+1}r_t
\end{equation}
Therefore, $R$ is the accumulated reward, and $r_t/t_f$ is the immediate reward for an individual period.

\subsubsection{Deterministic Policy Gradient}
\label{subsubsec:Deterministic Policy Gradient}

The policy $\pi$ is the strategy that the agent employs to determine the next action based on the current state. It maps states to actions, the actions that promise the highest reward. Mathematically, the policy can be represented by $\pi \colon S \rightarrow A$. By full exploitation, the policy can deterministically produce an action from a state. The optimal policy is obtained using a gradient ascent algorithm. To achieve this, a policy is specified by a set of parameters $\theta$, and $\boldsymbol{a_t} = \pi_{\boldsymbol{\theta}}(\boldsymbol{s_t})$. The performance matrix of $\pi_{\boldsymbol{\theta}}$ for the time interval $[0,t_f]$ is defined as the corresponding reward function of this interval,
\begin{equation}
    J_{[0,t_f]}(\pi_{\boldsymbol{\theta}}) = R(\boldsymbol{s_1},\pi_{\boldsymbol{\theta}}(\boldsymbol{s_1}),\dots,\boldsymbol{s_{t_f}},\pi_{\boldsymbol{\theta}}(\boldsymbol{s_f}),\boldsymbol{s_{t_f+1}})
\end{equation}

The initial parameter is initialized by randomization, and will be continuously updated along the gradient direction with a learning rate $\lambda$,
\begin{equation}
    \theta \rightarrow \theta + \lambda \triangledown_{\boldsymbol{\theta}}J_{[0,t_f]}(\pi_{\boldsymbol{\theta}})
\end{equation}

As a usual practice in reinforcement learning, the policy is updated upon mini-batches instead of the whole time frame as above. The true updating rule will then be, for each mini-batch $[t_{b_1},t_{b_2}]$, do 
\begin{equation}
        \theta \rightarrow \theta + \lambda \triangledown_{\boldsymbol{\theta}}J_{[t_{b_1},t_{b_2}]}(\pi_{\boldsymbol{\theta}})
\end{equation}

\section{Policy Networks}
\label{sec:Policy Networks}

Neural networks are function approximators, which are particularly useful in reinforcement learning when the state space or action space are too large to be completely known. Typically, it is used to approximate the policy function in this framework. 

The policy function $\pi_{\boldsymbol{\theta}}$ is constructed using three different neural networks, namely CNN, LSTM and RNN. Three important innovations are included in the framework that are crucial for the completeness and effectiveness. They are network topologies, the portfolio-vector memory and the stochastic mini-batch online learning scheme.

\subsection{Network Topologies}
\label{subsec:Network Topologies}

\begin{figure}[htbp]
    \centering
    \includegraphics[width=\textwidth]{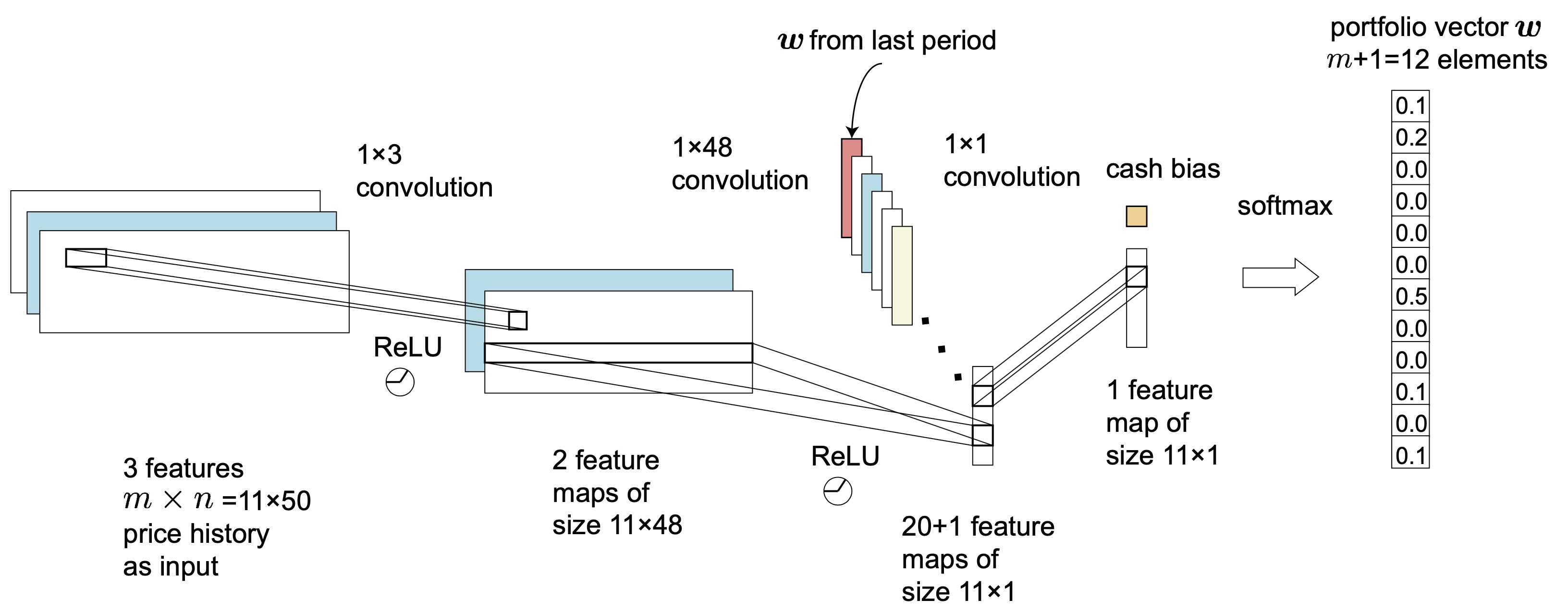}
    \caption{CNN Implementation of the EIIE}
    \label{fig:CNN}
\end{figure}

\begin{figure}[htbp]
    \centering
    \includegraphics[width=\textwidth]{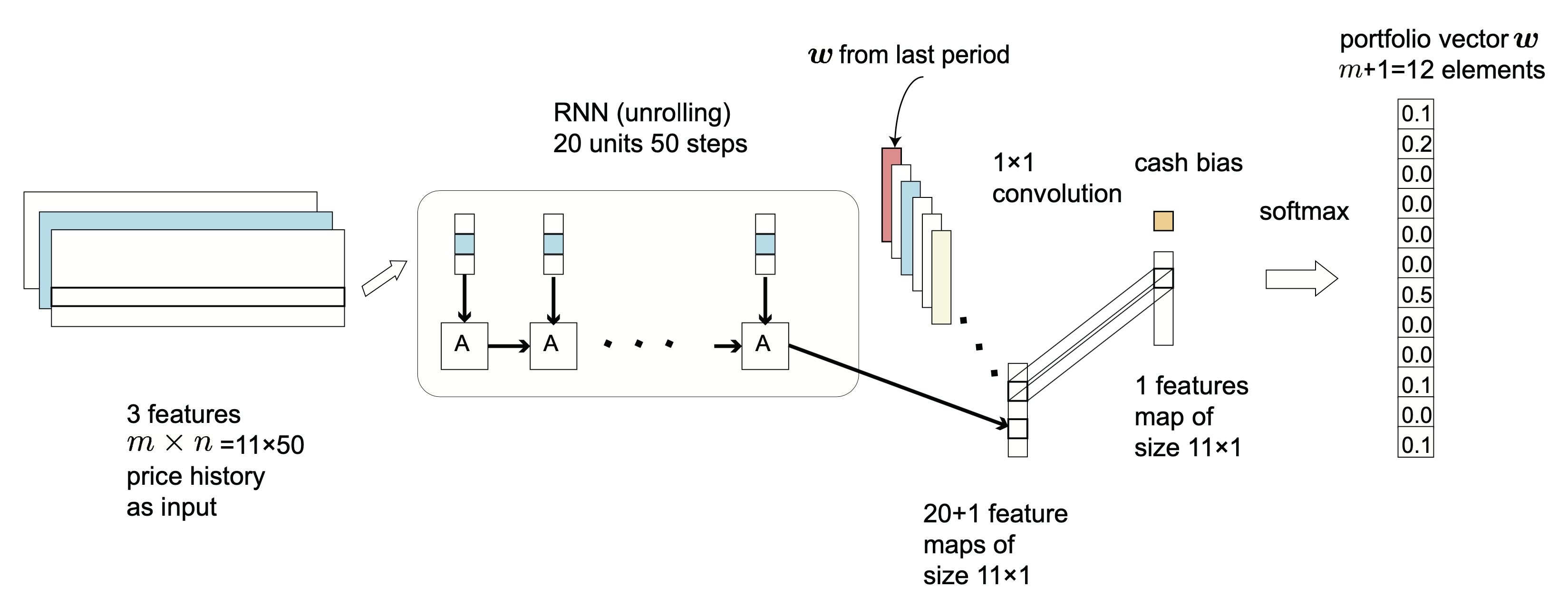}
    \caption{RNN (or LSTM) Implementation of the EIIE}
    \label{fig:RNN}
\end{figure}

The implementations of policy networks using CNN, RNN and LSTM are visualized using the figure extracted from the original paper, see Figure \ref{fig:CNN} and Figure \ref{fig:RNN}. For all implementations, the input is the price tensor $\boldsymbol{X_t}$ as well as the weight from last period $\boldsymbol{w_{t-1}}$. The output is the portfolio vector of the next period $\boldsymbol{w_{t}}$.

As one can see from both figures, the last hidden layers are the voting scores for all non-cash assets. The soft-max outcomes of these scores and a cash bias become the actual corresponding portfolio weights.  Please keep in mind that $\boldsymbol{w_{t-1}}$ is fed into the network just before the voting layer, differing from $\boldsymbol{X_t}$ being fed into the first layer.

For the CNN implementation, The first dimensions of all the local receptive fields in all feature maps are 1, making all rows isolated from each other until the soft-max activation. Apart from weight-sharing among receptive fields in a feature map, which is a usual CNN characteristic, parameters are also shared between rows in an EIIE configuration. Each row of the entire network is assigned with a particular asset, and is responsible to submit a voting score to the soft-max on the growing potential of the asset in the coming trading period. The input to the network is a $(3 \times m \times n)$ price tensor, comprising the highest, closing, and lowest prices of m non-cash assets over the past n periods. The outputs are the new portfolio weights. The previous portfolio weights are inserted as an extra feature map before the scoring layer, for the agent to minimize transaction costs.

The RNN implementation is the same as that of LSTM, with the only difference being RNN replacing LSTM in the first layer. It is a recurrent realization the Ensemble of Identical Independent Evaluators (EIIE). In this version, the price inputs of individual assets are taken by small recurrent subnets. These subnets are identical LSTMs or RNNs. 

An important feature in this framework is that the network will evaluate each of the $m$ assets independently but the parameters are shared among those streams. These streams are like independent but identical networks of smaller scopes, separately observing and assessing individual non-cash assets. They only interconnect at the softmax function, just to make sure their outputting weights are non-negative and summing up to unity. This special architecture of Identical Independent Evaluators (IIE) ensembled together to build up the mega topology is the reason that the authors named it as EIIE.

\subsection{Portfolio-Vector Memory}
\label{subsec:Portfolio-Vector Memory}
Inspired by the idea of experience replay memory \citep{mnih2016asynchronous}, the author introduced a dedicated Portfolio-Vector Memory (PVM) to store the network outputs, i.e. the weights in each step. The PVM is a stack of portfolios in chronological order. Before any network training, the PVM is initialized with uniform weights. In each training step, a policy network loads the portfolio vector of the previous period from the memory location at $t-1$, and overwrites the memory at $t$ with its output. As the parameters of the policy networks converge through many training epochs, the values in the memory also converge.

The advantages that PVM brings to the framework are significant:
\begin{enumerate}
    \item PVM allows the network to be trained simultaneously against data points within mini-batches, enormously improving training efficiency. 
    \item In the case of RNN versions of the networks, inserting previous outputs after the recurrent blocks avoids passing the gradients back to the deep RNN structures, circumventing the gradient vanishing problem.
\end{enumerate}

\subsection{Online Stochastic Batch Learning}
\label{subsec:Online Stochastic Batch Learning}

When this framework is applied to real-time trading, it is essential for the agent to update its policy when new data comes. The ever-ongoing nature of financial markets motivates the inclusion of online training. In this framework, the author proposed the so called Online Stochastic Batch Learning (OSBL) scheme.

At time $t$, the price movement information of period $t$ is available and added to the training set. After the agent takes trading orders at the beginning of period $t+1$, the policy network will be trained against $N_b$ randomly chosen mini-batches from this set. Each batch will be starting from $t_b$ with size $n_b$, The starting period is chosen from a geometric distribution, i.e. $t_b \sim GEO(\beta)$ or the probability $P_{\beta}(t_b)$ is,
\begin{equation}
    P_{\beta}(t_b)=\beta (1-\beta)^{t-t_b-n_b},\ t_b\leq t- n_b
\end{equation}
where $\beta \in (0,1]$ is the decay rate reflecting your beliefs on the importance of recent market events and $n_b$ is the number of periods in each mini-batch.
\newcommand{\Var}{\mathrm{Var}}
\chapter{Numerical Experiments and Evaluation}
\label{sec:Evaluation}

The deep reinforcement learning framework is examined and evaluated with two experiments. One is in the cryptocurrency market as it is in the original paper, the other one is in the stock market. We use a dataset different from the original paper in the first experiment and then totally change the asset to be selected from the equity market in the second experiment. 

We first explain the performance measures that are used to evaluate the performance of a particular portfolio selection strategy. Then the results of the two experiments are discussed.

\subsubsection{Performance Measures}
\label{subsubsec:Performance Measures}  

The following measures and factors will be used in our later discussions:
\begin{enumerate}
    \item Final Accumulated Portfolio Value (fAPV): the accumulated portfolio value over the whole time span of the back-test, with the initial value being 1. 
    \item Sharpe ratio: risk adjusted mean return, defined as the difference between the return of the investment and the risk-free return, divided by the standard deviation of the investment.. 
    \[
    SP=\frac{\mathop{\mathbb{E}}_t[\rho_t-\rho_F]}{\sqrt{\Var_t(\rho_t-\rho_F)}}\]
    where $\rho_t$ are the periodic returns defined in Equation \eqref{eq:rho_w_transaction} and $\rho_F$ is the the rate of return of a risk-free asset.
    \item Maximum Draw-down (MDD): the biggest loss from a peak to a trough,  mathematically
    \[
    D=\max_t\left(\max_{\tau > t}\frac{p_t-p_{\tau}}{p_t}\right)
    \]
    where $t \in (0,T]$
\end{enumerate}

\section{In the Cryptocurrency Market}
\label{sec:evalOne}  

Below is the time range for \textbf{Experiment 1}. All times below are in Coordinated Universal Time (UTC). All training sets start at $0$ o’clock. All price data are accessed through Poloniex’s official Application Programming Interface (API)\footnote{https://poloniex.com/support/api/}. Back-testing time range is the same as the testing time range, the only difference is that the former is online after training is complete  and the latter is offline during the training process.

\begin{itemize}
    \item \textbf{Experiment 1} Time Range: 2015-07-01 to 2017-07-01
    \begin{itemize}
        \item Training Time Range: 2015-07-01 to 2017-05-03
        \item Best-testing (Testing) Time Range: 2017-05-04 to 2017-07-01
    \end{itemize}
\end{itemize}

\begin{figure}[htbp]
    \centering
    \includegraphics[width=\textwidth]{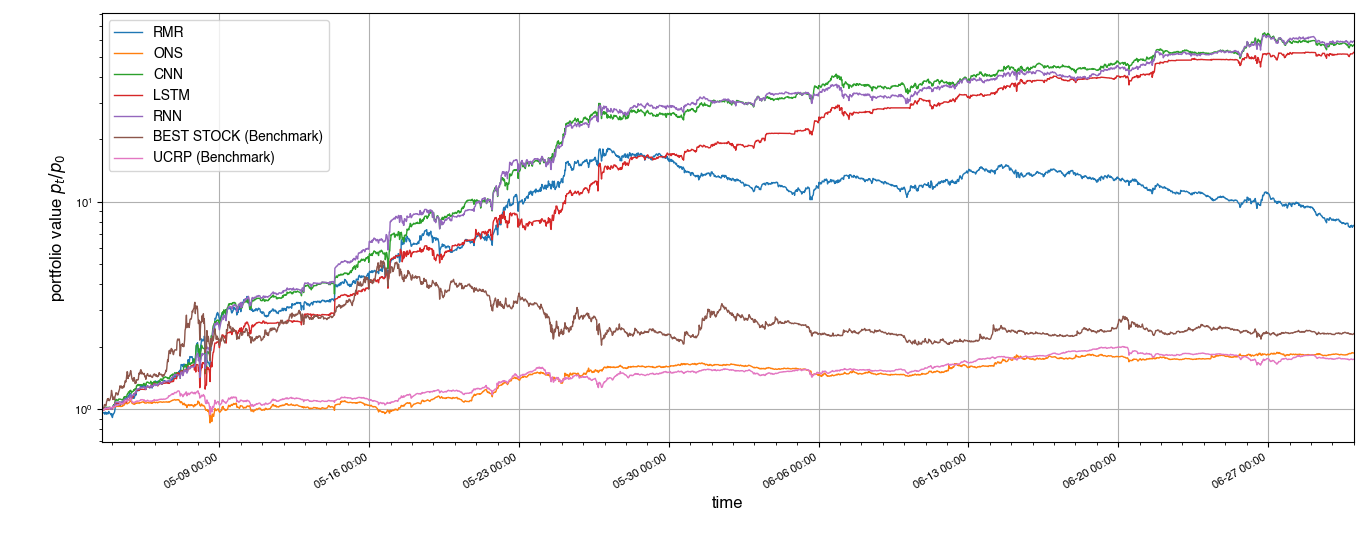}
    \caption{Log-scale accumulated wealth from 2017-05-04 to 2017-07-01 - \textit{Back-test 1}}
    \label{fig:Backtest1}
\end{figure}

More configurations for \textbf{Experiment 1} can be found in Appendix \ref{app:Experiment 1}, which lists all the hyper-parameters' values.

Figure \ref{fig:Backtest1} plots the APV against time in the back-test respectively for the CNN, RNN and LSTM EIIE networks, two selected benchmarks and two model-based strategies. The benchmarks Best Stock and UCRP are two good representatives of the market. We can see all EIIEs beat the market throughout the entirety of the back-tests.

\begin{table}[htbp]

  \centering

  \setlength\tabcolsep{3pt} 
  \footnotesize
  \resizebox{\textwidth}{!}{

    \begin{tabular}{c|ccc|cc|c}
    \toprule
          & \multicolumn{3}{c|}{\textbf{Test}} & \multicolumn{2}{c|}{\textbf{Backtest}} &  \\
    \textit{Network} & \textit{fAPV} & \textit{Log Mean} & \textit{Log Mean Free} & \textit{fAPV} & \textit{Log Mean} & \textit{Training Time(s)} \\
    \midrule
    CNN   & 8.938228607 & 0.000789027 & 0.003126783 & 56.98793309 & 0.001456354 & 590 \\
    RNN   & 25.48135948 & 0.001166407 & 0.002652416 & 59.49014754 & 0.001471834 & 1203 \\
    LSTM  & 42.89720917 & 0.001354037 & 0.002863562 & 52.48585405 & 0.001426709 & 3751 \\
    \bottomrule
    \end{tabular}%
    }
 \caption{Training and Back-testing results for CNN, RNN and LSTM - \textit{Back-test 1}}
  \label{tab:BT1table1}%
\end{table}%

Table \ref{tab:BT1table1} is the internal comparison between the three different EIIEs with different neural networks. One can observe a great reduction in log mean when including the transaction costs, which restates the importance of considering transaction costs in the real financial world. For all three networks, the test fAPV is less than that in back-testing, indicating the power of online learning. Among them, LSTM performs the best in terms of testing fAPV, while RNN is best with respect to back-testing fAPV. The out-performance of RNN over CNN and LSTM in back-testing is not significant though.

\begin{table}[htbp]

  \centering

  \setlength\tabcolsep{3pt} 
  \footnotesize
\resizebox{\textwidth}{!}{
    \begin{tabular}{llllllllll}
    \toprule
    \textbf{ Algorithms } & \textbf{ MDD } & \textbf{ fAPV } & \textbf{ SR } & \textbf{ - Days } & \textbf{ - Periods } & \textbf{ - Weeks } & \textbf{ + Days } & \textbf{ + Periods } & \textbf{ + Weeks } \\
    \midrule
     CNN  & 0.279 & 56.988 & 0.084 &                  659  &               1,379  &                    13  &               2,117  &               1,397  &               2,763  \\
     RNN  & 0.232 & \cellcolor[rgb]{ 1,  .78,  .808}\textcolor[rgb]{ .612,  0,  .024}{59.490} & \cellcolor[rgb]{ 1,  .78,  .808}\textcolor[rgb]{ .612,  0,  .024}{0.088} &                  679  &               1,373  &                    17  &               2,097  &               1,403  &               2,759  \\
     LSTM  & 0.255 & 52.486 & 0.083 & \cellcolor[rgb]{ 1,  .78,  .808}\textcolor[rgb]{ .612,  0,  .024}{                 514 } &               1,551  & \cellcolor[rgb]{ 1,  .78,  .808}\textcolor[rgb]{ .612,  0,  .024}{                   10 } & \cellcolor[rgb]{ 1,  .78,  .808}\textcolor[rgb]{ .612,  0,  .024}{              2,262 } &               1,187  & \cellcolor[rgb]{ 1,  .78,  .808}\textcolor[rgb]{ .612,  0,  .024}{              2,766 } \\
    \midrule
     Best Stock  & 0.608 & 2.310 & 0.025 &               1,403  &               1,432  &               1,163  &               1,373  &               1,328  &               1,613  \\
     UCRP  & 0.234 & 1.739 & 0.031 &               1,055  & \cellcolor[rgb]{ 1,  .78,  .808}\textcolor[rgb]{ .612,  0,  .024}{              1,264 } &                  650  &               1,721  & \cellcolor[rgb]{ 1,  .78,  .808}\textcolor[rgb]{ .612,  0,  .024}{              1,512 } &               2,126  \\
     UBAH  & 0.268 & 1.457 & 0.020 &               1,119  &               1,282  &                  890  &               1,657  &               1,494  &               1,886  \\
    \midrule
     ANTICOR  & \cellcolor[rgb]{ 1,  .78,  .808}\textcolor[rgb]{ .612,  0,  .024}{0.187} & 4.892 & 0.054 &                  870  &               1,327  &                  409  &               1,906  &               1,449  &               2,367  \\
     OLMAR  & 0.608 & 4.320 & 0.036 &               1,342  &               1,451  &               1,220  &               1,434  &               1,319  &               1,556  \\
     PAMR  & 0.980 & 0.041 & -0.048 &               2,018  &               1,589  &               2,030  &                  758  &               1,186  &                  746  \\
     WMAMR  & 0.563 & 1.821 & 0.021 &               1,409  &               1,486  &               1,166  &               1,367  &               1,287  &               1,610  \\
     CWMR  & 0.986 & 0.028 & -0.054 &               2,049  &               1,607  &               2,051  &                  727  &               1,168  &                  725  \\
     RMR  & 0.584 & 7.676 & 0.046 &               1,233  &               1,451  &               1,027  &               1,543  &               1,318  &               1,749  \\
    \midrule
     ONS  & 0.229 & 1.864 & 0.035 &               1,133  &               1,366  &                  837  &               1,643  &               1,410  &               1,939  \\
     UP   & 0.235 & 1.718 & 0.031 &               1,068  &               1,267  &                  649  &               1,708  &               1,509  &               2,127  \\
     EG   & 0.235 & 1.726 & 0.031 &               1,065  &               1,269  &                  648  &               1,711  &               1,507  &               2,128  \\
    \midrule
     BK   & 0.782 & 0.574 & -0.003 &               1,701  &               1,403  &               1,830  &               1,075  &               1,373  &                  946  \\
     CORNK  & 0.978 & 0.027 & -0.085 &               2,359  &               1,573  &               2,729  &                  417  &               1,203  &                    47  \\
     M0   & 0.289 & 1.890 & 0.028 &               1,156  &               1,288  &                  639  &               1,620  &               1,488  &               2,137  \\
    \bottomrule
    \end{tabular}%
}
 \caption{Performance of the three EIIEs, compared with traditional online selection algorithms - \textit{Back-test 1}}
  \label{tab:BT1table2}%
\end{table}%

As one can see from Table \ref{tab:BT1table2}, all three EIIEs outperform the other algorithms in terms of fAPV and SR. The Anticor algorithm has the lowest MDD. By including a fixed commission rate, many traditional algorithms perform terribly with fAPV even less than the initial value 1. We also include the counting of negative and positive periods, days and weeks, whose result shows the continuous profitability of the three EIIEs. The only situation when the reinforcement learning algorithms lose is when counting negative/positive periods but the scoring gap is small.

\section{In the Stock Market}
\label{sec:evalTwo}

In \textbf{Experiment 2}, the asset is selected from the stock market. Data is extracted through the Kibot’s official Application Programming Interface (API)\footnote{http://www.kibot.com/api/historical\_data\_api\_sdk.aspx}. The time range is set to be the past two years.

\begin{itemize}
    \item \textbf{Experiment 2} Time Range: 2017-12-12 to 2019-12-11
    \begin{itemize}
        \item Training Time Range: 2017-12-12 to 2019-10-13
        \item Back-testing (Testing) Time Range: 2019-10-14 to 2019-12-11
    \end{itemize}
\end{itemize}

\begin{figure}[htbp]
    \centering
    \includegraphics[width=\textwidth]{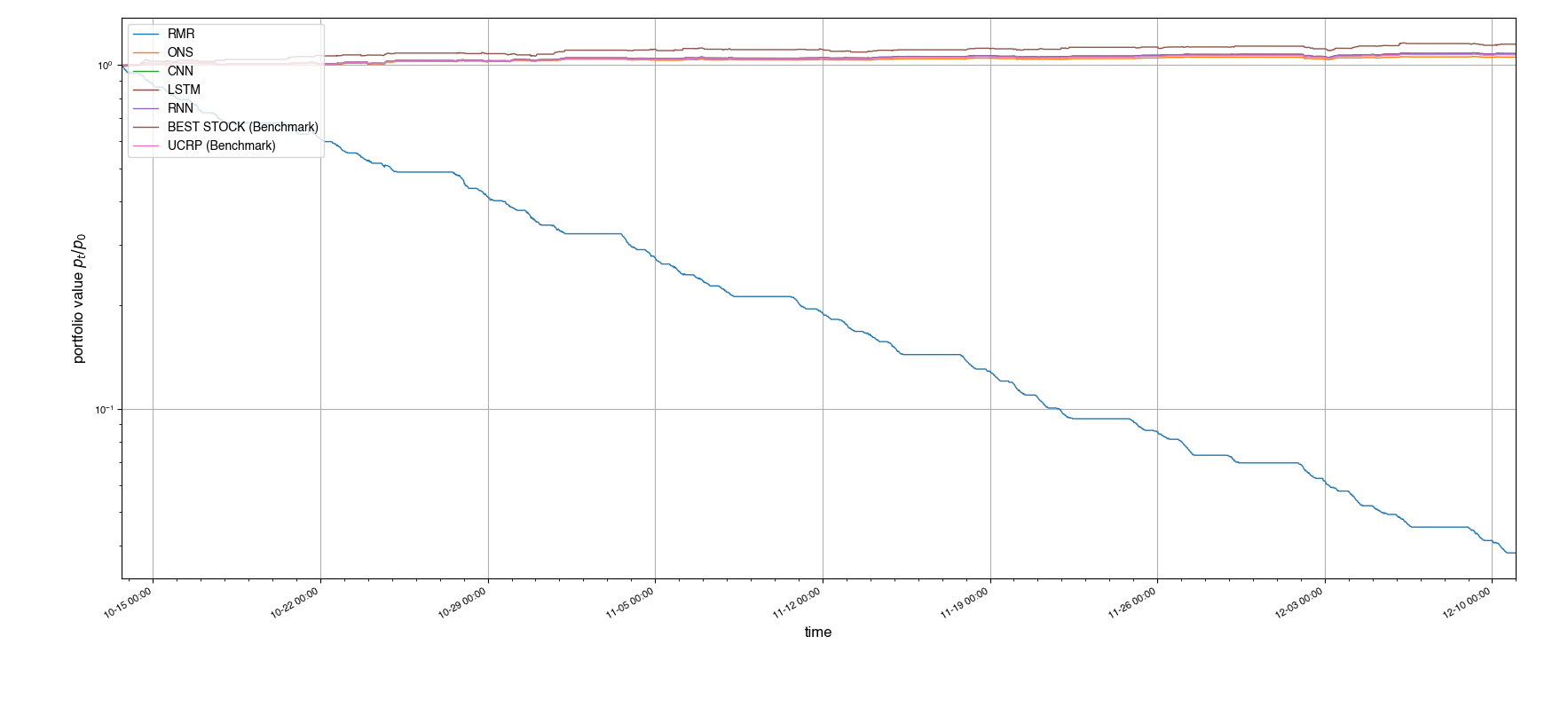}
    \caption{Log-scale accumulated wealth from 2019-10-14 to 2019-12-11 - \textit{Back-test 2}}
    \label{fig:Backtest2fig1}
\end{figure}

\begin{figure}[htbp]
    \centering
    \includegraphics[width=\textwidth]{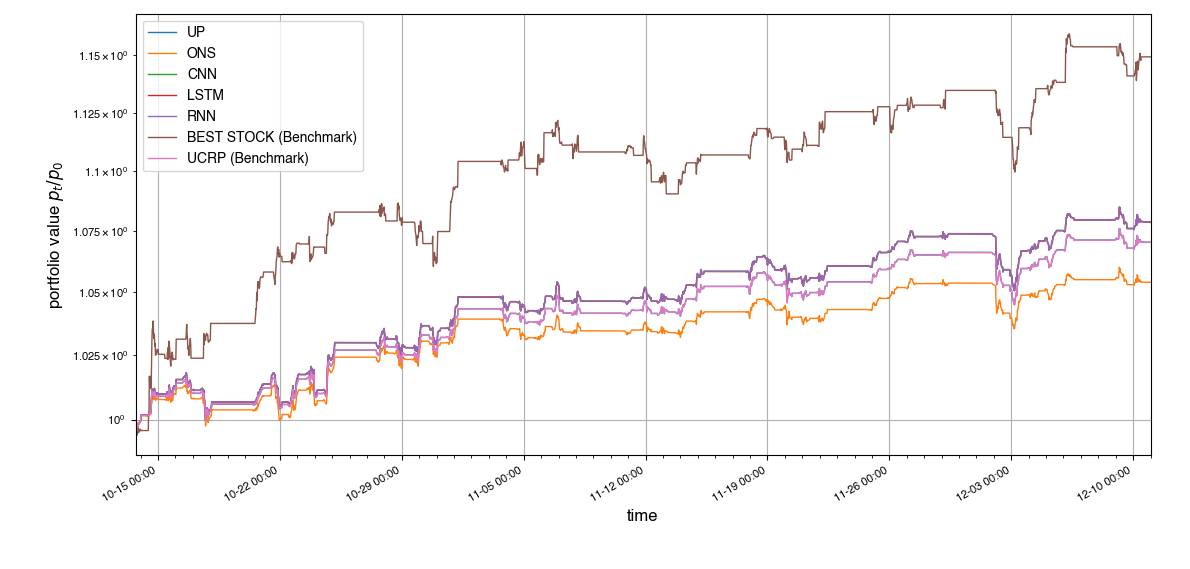}
    \caption{Log-scale accumulated wealth (excluding Follow-the-Loser) from 2019-10-14 to 2019-12-11 - \textit{Back-test 2}}
    \label{fig:Backtest2fig2}
\end{figure}

More configurations for \textbf{Experiment 2} can be found in Appendix \ref{app:Experiment 2}, which lists all the hyper-parameters' values.

Figure \ref{fig:Backtest2fig1} and Figure \ref{fig:Backtest2fig2} plot the APV against time in the three back-tests respectively for the CNN, RNN and LSTM EIIE networks, two selected benchmarks and two model-based strategies. Figure \ref{fig:Backtest2fig1} includes a type of Follow-the-Loser algorithm (RMR) that performs terrible in this case. Other Follow-the-Loser algorithms are also tested and similar outcomes are achieved, hence omitted plotting. As we know, Follow-the-Loser algorithm assumes that the under-performing assets will revert and outperform others in the subsequent periods. The terrible performance by the Follow-the-Loser algorithms may indicate the out-performing stocks remain strong in the time range and the under-performing stocks show less reversion behavior. This explanation is also validated by Figure \ref{fig:Backtest2fig2}, where the Best Stock algorithm performs the best and outperforms all others significantly. We know that the Best Stock algorithm is a special Buy and Hold algorithm that puts all capital on the stock with best performance in hindsight. The bet on that best stock being always the winner wins this time. This leaves the room for future developments. We may increase the testing set portion or adjust the decay rate $\beta$ to assign more importance to recent market event. Besides, we need more investigations in the stock selection process before training, as it is purely based on the ranking of the average trading volume.

\begin{table}[htbp]

  \centering

  \setlength\tabcolsep{3pt} 
  \footnotesize
  \resizebox{\textwidth}{!}{

     \begin{tabular}{c|ccc|cc|c}
    \toprule
          & \multicolumn{3}{c|}{\textbf{Test}} & \multicolumn{2}{c|}{\textbf{Backtest}} &  \\
    \textit{Network} & \textit{fAPV} & \textit{Log Mean} & \textit{Log Mean Free} & \textit{fAPV} & \textit{Log Mean} & \textit{Training Time(s)} \\
    \midrule
    CNN   & 1.081315756 & 2.82504E-05 & 2.85495E-05 & 1.078906536 & 2.74374E-05 & 613 \\
    LSTM  & 1.081547618 & 2.83314E-05 & 2.86326E-05 & 1.078754127 & 2.73863E-05 & 4629 \\
    RNN   & 1.081599355 & 2.83488E-05 & 2.86426E-05 & 1.078838147 & 2.74136E-05 & 1905 \\
    \bottomrule
    \end{tabular}%
    }
 \caption{Training and Back-testing results for CNN, RNN and LSTM - \textit{Back-test 2}}
  \label{tab:BT2table1}%
\end{table}%

Table \ref{tab:BT2table1} is the internal comparison between the three different EIIEs with different neural networks. Interesting things happen here. We notice that the performance measures in the table are close to each other among the three networks, except the training time which is determined by the nature of the underlying network. The log mean doesn't decrease as much as that in Table \ref{tab:BT1table1}. Upon further investigation into the back-testing, we find that a nearly uniform portfolio is kept in the whole back-testing process. The weight for each non-cash asset is about $\frac{1}{m}$. In this case, all three EIIEs perform similarly and can be seen as the Uniform Constant Rebalanced Portfolio (UCRP), which is validated by the nearly same trends in Figure \ref{fig:Backtest2fig2}. The minor influence of transaction costs on log mean can also be explained by the less frequent rebalancing. This result sheds light into what we have learnt in ActSc 972, where we have discussed about how efficient the $\frac{1}{N}$ strategy is. \cite{demiguel2007optimal} compared the performance of many different portfolio selection methods to the equal asset weights. They found the equally weighted strategy outperformed the other strategies out‐of‐sample on a variety of financial datasets. Even this naive algorithm is not the best but the Best Stock algorithm, it still ranks top among other algorithms. The reinforcement learning process gradually makes our agent follow this equally weighted strategy.

\begin{table}[htbp]

  \centering

  \setlength\tabcolsep{3pt} 
  \footnotesize
\resizebox{\textwidth}{!}{
    \begin{tabular}{llllllllll}
    \toprule
    \textbf{ Algorithms } & \textbf{ MDD } & \textbf{ fAPV } & \textbf{ SR } & \textbf{ - Days } & \textbf{ - Periods } & \textbf{ - Weeks } & \textbf{ + Days } & \textbf{ + Periods } & \textbf{ + Weeks } \\
    \midrule
     CNN  & 0.023 & 1.079 & \cellcolor[rgb]{ 1,  .78,  .808}\textcolor[rgb]{ .612,  0,  .024}{0.042} &                  659  &               1,387  &               1,645  &               1,381  & \cellcolor[rgb]{ 1,  .78,  .808}\textcolor[rgb]{ .612,  0,  .024}{                 703 } & \cellcolor[rgb]{ 1,  .78,  .808}\textcolor[rgb]{ .612,  0,  .024}{              2,364 } \\
     RNN  & 0.023 & 1.079 & 0.042 &                  679  &               1,388  &               1,690  &               1,380  &                  690  & \cellcolor[rgb]{ 1,  .78,  .808}\textcolor[rgb]{ .612,  0,  .024}{              2,364 } \\
     LSTM  & 0.023 & 1.079 & 0.042 & \cellcolor[rgb]{ 1,  .78,  .808}\textcolor[rgb]{ .612,  0,  .024}{                 514 } &               1,388  &               1,784  &               1,380  &                  682  & \cellcolor[rgb]{ 1,  .78,  .808}\textcolor[rgb]{ .612,  0,  .024}{              2,364 } \\
    \midrule
     Best Stock  & 0.031 & \cellcolor[rgb]{ 1,  .78,  .808}\textcolor[rgb]{ .612,  0,  .024}{1.149} & 0.042 &               1,403  &                  851  & \cellcolor[rgb]{ 1,  .78,  .808}\textcolor[rgb]{ .612,  0,  .024}{                 319 } &               1,278  &                  367  &               2,349  \\
     UCRP  & 0.021 & 1.070 & 0.041 &               1,055  &                  854  &                  649  &               1,379  &                  663  &               2,343  \\
     UBAH  & 0.021 & 1.072 & 0.042 &               1,119  & \cellcolor[rgb]{ 1,  .78,  .808}\textcolor[rgb]{ .612,  0,  .024}{                 840 } &                  578  & \cellcolor[rgb]{ 1,  .78,  .808}\textcolor[rgb]{ .612,  0,  .024}{              1,384 } &                  663  &               2,345  \\
    \midrule
     ANTICOR  & 0.442 & 0.560 & -0.243 &                  870  &               2,345  &               1,674  &                  105  &                  363  &                    10  \\
     OLMAR  & 0.960 & 0.040 & -0.443 &               1,342  &               2,221  &                  837  &                      4  &                  216  &                     0    \\
     PAMR  & 0.956 & 0.044 & -0.515 &               2,018  &               2,224  &               1,110  &                      5  &                  132  &                     0    \\
     WMAMR  & 0.673 & 0.326 & -0.278 &               1,409  &               2,238  &               1,105  &                    15  &                  273  &                     0    \\
    \midrule
     ONS  & 0.019 & 1.054 & 0.033 &               1,133  &                  906  &                  674  &               1,327  &                  640  &               2,222  \\
     UP   & 0.021 & 1.071 & 0.041 &               1,068  &                  854  &                  649  &               1,379  &                  663  &               2,343  \\
     EG   & 0.021 & 1.070 & 0.041 &               1,065  &                  854  &                  649  &               1,379  &                  663  &               2,343  \\
    \midrule
     M0   & \cellcolor[rgb]{ 1,  .78,  .808}\textcolor[rgb]{ .612,  0,  .024}{0.013} & 1.021 & 0.023 &               1,156  &               1,396  &               2,125  &               1,372  &                  643  &               1,950  \\
    \bottomrule
    \end{tabular}%
}
 \caption{Performance of the three EIIEs, compared with traditional online selection algorithms - \textit{Back-test 2}}
  \label{tab:BT2table2}%
\end{table}%

Table \ref{tab:BT2table2}, once again, validates our explanation above. The three EIIEs perform similarly to UCRP and even a little bit better in terms of fAPV. Even though they are not as good as the Best Stock algorithm, the gap in between is tiny. Follow-the-Loser algorithms have the worst performance of all\footnote{some of the Follow-the-Loser algorithms can not been implemented because the $\Sigma$ matrix is singular}.

\chapter{Conclusion and Future Work}
\label{sec:Conclusion}

The original paper proposed a deep reinforcement learning framework to solve the financial portfolio management problem. The core architecture is the EIIE meta topology, which is able to accommodate many types of weight-sharing neural network structures in the lower level. Additional innovations include the use of Portfolio Vector Memory (PVM) and the implementation of the Online Stochastic Batch Learning (OSBL) scheme. The former stores portfolio vectors in a separate memory such that transaction costs can be taken into account, the agent can avoid oversized reallocations between consecutive actions and the framework can be trained in parallel within batching. The latter governs the online learning process so that the agent can continuously digest constant incoming market information while trading.

The framework is implemented, trained and back-tested in two different markets, i.e. the cryptocurrency market and the stock market. In all experiments, the framework was realized using three different underlining networks, a CNN, a RNN and a LSTM. The profitability of all surpasses traditional portfolio-selection methods, as demonstrated in the paper by the outcomes of \textbf{Experiment 1}. Among the three EIIE networks, LSTM has lower scores than that of CNN and RNN. The gap in performance between the two RNN species under the same framework might be explained by the belief that history repeats itself. Not being designed to forget its input history, a vanilla RNN is more able than a LSTM to exploit repetitive patterns in price movements for higher yields. The gap might also be due to lack of fine-tuning in hyper-parameters for the LSTM. In the experiments, the same set of structural hyper-parameters was used for both basic RNN and LSTM.

Surprisingly, when it comes to the stock market, the magic of this framework disappeared and fell back to be equivalent to the Uniformly Constant Rebalanced Portfolios algorithm or the equally weighted strategy. The reinforcement learning gradually learns to sell high and buy low in order to take advantage of any mean  reversion, even though this reversion is not significant in back-testing.

The future work that can be done to further examine or improve the framework is listed as follows:

\begin{enumerate}
    \item Modify the framework to deal with situations when the two assumption in Section \ref{sec:Two assumptions} are violated.
    \item Try other reward functions in order to include awareness of longer-term market reactions.
    \item Include technical indicators in the input tensor, e.g. exponential moving average (EMA), volume weighted average price (VWAP), relative strength index (RSI), etc.
    \item Hyper-parameter tuning for stock marketS to see if the performance can be improved.
    \item Select assets more carefully rather than just depending on the average trading volume. 
\end{enumerate}

\clearpage
\bibliographystyle{dcu}
\renewcommand*{\bibname}{Bibliography}

\addcontentsline{toc}{chapter}{\textbf{Bibliography}}

\bibliography{refs}

\nocite{*}


\appendix
\chapter*{APPENDICES}
\addcontentsline{toc}{chapter}{Appendices}
\colorlet{punct}{red!60!black}
\definecolor{background}{HTML}{EEEEEE}
\definecolor{delim}{RGB}{20,105,176}
\colorlet{numb}{magenta!60!black}

\lstdefinelanguage{json}{
    basicstyle=\normalfont\ttfamily,
    numbers=left,
    numberstyle=\scriptsize,
    stepnumber=1,
    numbersep=8pt,
    showstringspaces=false,
    breaklines=true,
    frame=lines,
    backgroundcolor=\color{background},
    literate=
     *{0}{{{\color{numb}0}}}{1}
      {1}{{{\color{numb}1}}}{1}
      {2}{{{\color{numb}2}}}{1}
      {3}{{{\color{numb}3}}}{1}
      {4}{{{\color{numb}4}}}{1}
      {5}{{{\color{numb}5}}}{1}
      {6}{{{\color{numb}6}}}{1}
      {7}{{{\color{numb}7}}}{1}
      {8}{{{\color{numb}8}}}{1}
      {9}{{{\color{numb}9}}}{1}
      {:}{{{\color{punct}{:}}}}{1}
      {,}{{{\color{punct}{,}}}}{1}
      {\{}{{{\color{delim}{\{}}}}{1}
      {\}}{{{\color{delim}{\}}}}}{1}
      {[}{{{\color{delim}{[}}}}{1}
      {]}{{{\color{delim}{]}}}}{1},
}

\chapter{Proofs of Theorems}
\label{AppendixA}
\section{Proof of the Formula for Transaction Remainder Factor}

During the portfolio reallocation from $\boldsymbol{w_t'}$ to $\boldsymbol{w_t}$, some or all of the amount of asset $i$ needs to be sold, if $p_t'w_{t,i}' > p_t w_{t,i} ,\text{or}\ w_{t,i}' > \mu_t w_{t,i}$ The total amount of cash obtained by all selling is
\begin{equation}
 (1-c_s)p_t' \sum_{i=1}^{m}(w_{t,i}' - \mu_t w_{t,i})^+
\end{equation}
where $0\leq c_s < 1$ is the commission rate for selling, and $\boldsymbol{v}^+ = \text{ReLu}(\boldsymbol{v})$ is the element-wise rectified linear function. This money and the original cash reserve $p_t' w_{t,0}'$ taken away from the new reserve $\mu_t p_t' w_{t,0}'$ will be used to buy new assets,

\begin{equation}
     (1-c)\left[ w_{t,0}'+(1-c_s) \sum_{i=1}^{m}(w_{t,i}' - \mu_t w_{t,i})^+ - \mu_t w_{t,0} \right]=\sum_{i=1}^{m}(\mu_t w_{t,i}-w_{t,i}')^+
     \label{eq:A2}
\end{equation}
where $0\leq c_p < 1$ is the commission rate for purchasing, and $p_t'$ has been cancelled out on both sides. Using the identity $(a-b)^+-(b-a)^- = a-b$, and the fact that $w_{t,0}' + \sum_{i=1}^m w'_{i,t}=w_{t,0} + \sum_{i=1}^m w_{i,t}$, Equation \eqref{eq:A2} is simplified to
\begin{equation}
    \mu_t=\frac{1}{1-c_p w_{t,0}}\left[1-c_p w'_{t,0}-(c_s+c_p-c_s c_p)\sum_{i=1}^{m}(w'_{t,i}-\mu_t w_{t,i})^+ \right]
\end{equation}

$\mu_t$ can not be solved analytically, but can be solved recursively. The following is the numerical method to solve for $\mu_t$:

The sequence $\{\Tilde{\mu}_t^{(k)}\}$ is defined as 
\begin{equation}
    \left\{\Tilde{\mu}_t^{(k)}|  \Tilde{\mu}_t^{(0)} = \mu_0 \And \Tilde{\mu}_t^{(k)} = f(\Tilde{\mu}_t^{(k-1)}) \right\}
\end{equation}
where $f(\mu)$ is,
\begin{equation}
    f(\mu) \coloneqq \frac{1}{1-c_p w_{t,0}}\left[1-c_p w'_{t,0}-(c_s+c_p-c_s c_p)\sum_{i=1}^{m}(w'_{t,i}-\mu w_{t,i})^+ \right]
\end{equation}
and $\mu_0$ is the initial guess.

\chapter{Configurations for the Experiments}
\label{AppendixB}

\section{Experiment 1 Configurations}
\label{app:Experiment 1}

\begin{lstlisting}[language=json,firstnumber=1]
{
  "layers":
  [
    {"filter_shape": [1, 2], "filter_number": 3, "type": "ConvLayer"},
    {"filter_number":10, "type": "EIIE_Dense", "regularizer": "L2", "weight_decay": 5e-9},
    {"type": "EIIE_Output_WithW","regularizer": "L2", "weight_decay": 5e-8}
  ],
  "training":{
    "steps":80000,
    "learning_rate":0.00028,
    "batch_size":109,
    "buffer_biased":5e-5,
    "snap_shot":false,
    "fast_train":true,
    "training_method":"Adam",
    "loss_function":"loss_function6"
  },

  "input":{
    "window_size":31,
    "coin_number":11,
    "global_period":1800,
    "feature_number":3,
    "test_portion":0.08,
    "online":false,
    "start_date":"2015/07/01",
    "end_date":"2017/07/01",
    "volume_average_days":30,
    "portion_reversed": false
  },

  "trading":{
    "trading_consumption":0.0025,
    "rolling_training_steps":85,
    "learning_rate":0.00028,
    "buffer_biased":5e-5
  }
}

\end{lstlisting}

The above is the configuration for \textbf{Experiment 1} with CNN. LSTM (RNN: ``type'' change to ``EIIE\_RNN'') configuration only differs in the layer input as bellows:
\begin{lstlisting}[language=json,firstnumber=1]
{  "layers": [
        {
            "dropouts": null,
            "neuron_number": 20,
            "type": "EIIE_LSTM" 
        },
        {
            "regularizer": "L2",
            "type": "EIIE_Output_WithW",
            "weight_decay": 5e-08
        }
}
\end{lstlisting}

\section{Experiment 2 Configurations}
\label{app:Experiment 2}
\begin{lstlisting}[language=json,firstnumber=1]
{
  "layers":
    [
      {"filter_shape": [1, 2], "filter_number": 3, "type": "ConvLayer"},
      {"filter_number":10, "type": "EIIE_Dense", "regularizer": "L2", "weight_decay": 5e-9},
      {"type": "EIIE_Output_WithW","regularizer": "L2", "weight_decay": 5e-8}
    ],
  "training":{
    "steps":80000,
    "learning_rate":0.00028,
    "batch_size":109,
    "buffer_biased":5e-5,
    "snap_shot":false,
    "fast_train":true,
    "training_method":"Adam",
    "loss_function":"loss_function6"
  },

  "input":{
    "window_size":31,
    "coin_number":11,
    "global_period":1800,
    "feature_number":3,
    "test_portion":0.08,
    "online":false,
    "start_date":"2017/12/12",
      "end_date":"2019/12/11",
    "volume_average_days":30,
    "portion_reversed": false
  },

  "trading":{
    "trading_consumption":0.0025,
    "rolling_training_steps":85,
    "learning_rate":0.00028,
    "buffer_biased":5e-5
  }
}

\end{lstlisting}

The above is the configuration for \textbf{Experiment 2} with CNN. LSTM (RNN: ``type'' change to ``EIIE\_RNN'') configuration only differs in the layer input as follows:
\begin{lstlisting}[language=json,firstnumber=1]
{  "layers": [
        {
            "dropouts": null,
            "neuron_number": 20,
            "type": "EIIE_LSTM" 
        },
        {
            "regularizer": "L2",
            "type": "EIIE_Output_WithW",
            "weight_decay": 5e-08
        }
}
\end{lstlisting}

\chapter{GitHub Repository}
\label{AppendixC}
The implementation for this paper is stored at the repository on GitHub, \url{https://github.com/jackieli19/PGPortfolio/tree/Stock}. One may find all the codes, figures, tables and data there. Below is only the user guide. 

Our experiments are all trained and back-testing on MacBook Pro 2018 without using GPU.

\section{Quickstart}
\label{quickstart}

\begin{enumerate}
\def\labelenumi{\arabic{enumi}.}
\item
  Edit
  \href{pgportfolio/net_config.json}{\texttt{pgportfolio/net\_config.json}}
\item
  Generate an agent:
\end{enumerate}

\begin{verbatim}
python main.py --mode=generate --repeat=1
\end{verbatim}

\begin{enumerate}
\def\labelenumi{\arabic{enumi}.}
\setcounter{enumi}{2}
\item
  Download the data:
\end{enumerate}

\begin{verbatim}
python main.py --mode=download_data
\end{verbatim}

\begin{enumerate}
\def\labelenumi{\arabic{enumi}.}
\setcounter{enumi}{3}
\item
  Train the agent:
\end{enumerate}

\begin{verbatim}
python main.py --mode=train --processes=1
\end{verbatim}

\begin{enumerate}
\def\labelenumi{\arabic{enumi}.}
\setcounter{enumi}{4}
\item
  Compare the result with other algorithms:
\end{enumerate}

\begin{verbatim}
python main.py --mode=plot --algos=crp,1 --labels=crp,nnagent
python main.py --mode=table --algos=1,ons --labels=nntrader,ons
\end{verbatim}

See below for details on each step.

\section{Configuration File}
\label{configuration-file}

\texttt{pgportfolio/net\_config.json} contains all the configuration
parameters. The software can be configured by modifying this file and
without any changes to the code.

The parameters are classified into four categories: Network Topology, Market Data, Training and Trading.

\section{Training the agent}\label{training-the-agent}

In order to train the agent perform the following steps: 1.
\emph{(Optional)} Modify the configuration in
\texttt{pgportfolio/net\_config.json} according to your desired agent
configuration.

\begin{enumerate}
\def\labelenumi{\arabic{enumi}.}
\setcounter{enumi}{1}
\tightlist
\item
  From the main folder, run:
\end{enumerate}

\begin{verbatim}
python main.py --mode=generate --repeat=n
\end{verbatim}

where \texttt{n} is a positive integer indicating the number of replicas
you would like to train. This will create \texttt{n} subfolders in the
\texttt{train\_package} folder. Each subfolder contains a copy of the
\texttt{net\_config.json} file. The random seed of each subfolder
runs from \texttt{0} to \texttt{n-1}. \emph{\emph{Please note that
agents with different random seeds can have very different
performances.}}

\begin{enumerate}
\def\labelenumi{\arabic{enumi}.}
\setcounter{enumi}{2}
\tightlist
\item
  (Optional) Download the data with the command:
\end{enumerate}

\begin{verbatim}
python main.py --mode=download_data
\end{verbatim}

\begin{enumerate}
\def\labelenumi{\arabic{enumi}.}
\setcounter{enumi}{3}
\tightlist
\item
  Train your agents with the command:
\end{enumerate}

\begin{verbatim}
python main.py --mode=train --processes=1
\end{verbatim}

\begin{verbatim}
* This will start training the `n' agents one at a time. 
Do not start more than 1 processes if you want to download data 
online.
* `--processes=m' starts `m' parallel training processes
* `--device=gpu' can be added if your tensorflow supports GPU.
 - On _GTX1080Ti_ you should be able to run 4-5 training 
 processes simultaneously.
 - On _GTX1060_ you should be able to run 2-3 training
 processes simultaneously.
\end{verbatim}

\begin{enumerate}
\def\labelenumi{\arabic{enumi}.}
\setcounter{enumi}{4}
\item
  Each training run is composed of 2 phases: \textbf{Training} and
  \textbf{Backtest}.

  \begin{itemize}
  \tightlist
  \item
    During the \textbf{Training} phase, the agent is trained on the
    training fraction of the global data matrix. The log looks like
    this:
  \end{itemize}
\end{enumerate}

\begin{verbatim}
average time for data accessing is 0.0015480489730834962
average time for training is 0.009850282192230225
==============================
step 2000
------------------------------
the portfolio value on test set is 2.118205
log_mean is 0.00027037683
loss_value is -0.000270
log mean without commission fee is 0.000341
\end{verbatim}

\begin{itemize}
\tightlist
\item
  After training is completed, the \textbf{Backtest} phase begins. This
  uses a rolling training window, i.e.~it performs online learning in
  supervised learning. The log looks like this:
\end{itemize}

\begin{verbatim}
the step is 536
total assets are 4.314677 BTC
\end{verbatim}

\begin{enumerate}
\def\labelenumi{\arabic{enumi}.}
\setcounter{enumi}{5}
\item
  Once training and backtest are completed, you can check the result
  summary of the training in \texttt{train\_package/train\_summary.csv}
\item
  Tune the hyper-parameters based on the summary, and go to 1 again.
\end{enumerate}

\section{Training results}\label{training-results}

Once training is completed, each subfolder in \texttt{train\_package}
will contain several output artifacts:

\begin{itemize}
\item
  \texttt{programlog}: a log file generated during training. This
  contains the same information that was visualized in output during
  training and backtesting.
\item
  \texttt{tensorboard}: a folder containing the events for thensorboard.
  You can visualize its content by running tensorboard:
  e.g.~\texttt{tensorboard\ -\/-logdir=train\_package/1}.
\item
  \texttt{netfile.*}: the model checkpoints. These can be used to
  restore a previously trained model.
\item
  \texttt{train\_summary.csv}: a file with summary information like:
  network configuration, portfolio value on validation set and test set
  etc.
\end{itemize}

\section{Download Data}\label{download-data}

To prefetch data to the local database without starting a training run:

\begin{verbatim}
python main.py --mode=download_data
\end{verbatim}

The program will use the configurations in
\texttt{pgportfolio/net\_config.json} to select coins and download
necessary data to train the network. * Download speed could be very slow
and sometimes even have errors in China. * If you can cannot download
data, please check the first release where there is a \texttt{Data.db}
file. Copy the file into the database folder. Make sure the
\texttt{online} in \texttt{input} in \texttt{net\_config.json} is
\texttt{false} and run the example. Note that using the this file, you
shouldn't make any changes to the input data configuration (for example
\texttt{start\_date}, \texttt{end\_date} or \texttt{coin\_number})
otherwise the results may not be correct.

\section{Backtest}\label{backtest}

To execute a backtest with rolling training (i.e.~online learning in
supervised learning) on the target model run:

\begin{verbatim}
python main.py --mode=backtest --algo=1
\end{verbatim}

\begin{itemize}
\tightlist
\item
  \texttt{-\/-algo} could be either the name of a traditional method or
  the index of the training folder
\end{itemize}

\section{Plotting}\label{plotting}

To plot the results run:

\begin{verbatim}
python main.py --mode=plot --algos=crp,1 --labels=crp,nnagent
\end{verbatim}

\begin{itemize}
\tightlist
\item
  \texttt{-\/-algos}: comma separated list of traditional algorithms and
  agent indexes
\item
  \texttt{-\/-labels}: comma separated list of names that appear in the
  plot legend
\end{itemize}

\section{Table summary}\label{table-summary}

You can present a summary of the results typing:

\begin{verbatim}
python main.py --mode=table --algos=1,ons --labels=nntrader,ons
\end{verbatim}

\begin{itemize}
\tightlist
\item
  \texttt{-\/-algos} and \texttt{-\/-labels} are the same as in the plotting
  case. Labels indicate the row indexes.
\end{itemize}

\begin{itemize}
\tightlist
\item
  use \texttt{-\/-format} arguments to change the format of the table, which
  could be \texttt{raw} \texttt{html} \texttt{csv} or \texttt{latex}.
  The default one is raw.
\end{itemize}

\end{document}